\documentclass[12pt]{article}
\usepackage{amssymb}
\usepackage{amsmath}
\usepackage{subcaption}
\usepackage{booktabs}
\usepackage{lineno}
\usepackage{graphicx}
\usepackage{authblk}

\begin{document}

\title{A Monte-Carlo simulation of the equilibrium beam polarization in ultra-high energy electron~(positron) storage rings
\thanks{To be published in Nucl. Instrum. Meth. A.}}




\author{Zhe~Duan$^{a,b}$, Mei~Bai$^c$, Desmond~P.~Barber$^d$, Qing~Qin$^a$ \\
$^a$ \small{Key Laboratory of Particle Acceleration Physics and Technology, Institute of High Energy Physics, Chinese Academy of Sciences, 100049 Beijing, China} \\
$^b$ \small{University of Chinese Academy of Sciences, 100049 Beijing, China} \\
$^c$ \small{Forschungszentrum J{\"u}lich GmbH, 52428 J{\"u}lich, Germany} \\
$^d$ \small{Deutsches Elektronen-Synchrotron, DESY, 22607 Hamburg, Germany}
}
\allowdisplaybreaks
\maketitle

\begin{center}
\textbf{Abstract}
\end{center}

\small{
    With the recently emerging global
    interest in building a next generation of circular
    electron-positron colliders to study the properties of the Higgs
    boson, and other important topics in particle physics at ultra-high
    beam energies, it is also important to pursue the possibility of
    implementing polarized beams at this energy scale. It is
    therefore necessary to set up simulation tools to evaluate the
    beam polarization at these ultra-high beam energies. In this
    paper, a Monte-Carlo simulation of the equilibrium beam
    polarization based on the Polymorphic Tracking Code(PTC)
    (Schmidt et al., 2002 \cite{Schmidt:2002vp}) is described. 
    The  simulations are for a model storage ring with parameters similar to those of proposed circular colliders in this energy
range, and they are compared with the suggestion (Derbenev et al., 1978 \cite{Derbenev:1979tm}) that there are different regimes for the spin dynamics underlying the polarization of a beam
in the presence of synchrotron radiation at ultra-high beam energies.
     In particular, it has been
    suggested that the so-called ``correlated'' crossing of spin
    resonances during synchrotron oscillations at current energies,
    evolves into ``uncorrelated'' crossing of spin resonances at
    ultra-high energies. 
    }

\section*{Introduction}
With the discovery of the Higgs boson by the ATLAS and CMS experiments
at CERN's Large Hadron Collider(LHC), it becomes natural to measure
its properties as precisely as possible, at the LHC or future
electron-positron colliders. Besides the International Linear
Collider, there is an alternative possibility, namely to build a
circular electron-positron collider, with the merit of a higher
luminosity-to-cost ratio, and the potential that it be upgraded to a
proton-proton collider later. Currently, there are two major design
studies of such a circular electron positron collider in the world,
the CEPC \cite{CEPC} and the FCC-ee(TLEP) \cite{FCC-ee}. These
designs open the possibility of precision measurements at the Z pole,
at the WW threshold, at the HZ cross section maximum, and even at the
$t\bar{t}$ threshold, with an unprecedented accuracy. Then it is
natural to pursue the possibility of implementing polarized electron
and positron beams at these ultra-high beam energies, for precision
beam energy calibration with transverse beam polarization, or
colliding-beam experiments with longitudinal beam polarization.

The motion of the spin expectation value (the ``spin'') $\vec{S}$ of a
relativistic charged particle traveling in electromagnetic fields is
described by the Thomas-BMT equation \cite{L.H.Thomas1927,
  V.Bargmann1959},
\begin{eqnarray}
    \label{Thomas-BMT}
    \frac{d\vec{S}}{d\theta}&=&\vec{\Omega}(\vec{u}, \theta)\times\vec{S} \nonumber \\
    \vec{\Omega}(\vec{u}, \theta)&=&\vec{\Omega}_0(\theta)+\vec{\omega}(\vec{u}, \theta),
\end{eqnarray}
where $\theta$ is the azimuthal angle, $\vec{\Omega}_0$ is due to the
fields on the closed orbit, and $\vec{\omega}$ is due to the fields at
the beam coordinates $\vec{u}= (x, p_x, y, p_y, z, \delta)$ 
with respect to the closed orbit, where $x$, $y$ and $z$ represent the
horizontal, vertical and longitudinal displacements, $p_x$ and $p_y$ are
horizontal and vertical canonical momenta, and $\delta={\Delta}E/E_0$ is the
relative energy deviation. As we shall see, it is often necessary to describe spin motion with the help 
of a unit vector field
${\hat n}(\vec{u},\theta)$ (the ''invariant spin field'', or ISF for short) 
\cite{D.P.Barber2004}. This satisfies the Thomas-BMT equation along particle trajectories and it is
periodic: $\hat{n}(\vec{u};\theta)=\hat{n}(\vec{u};\theta+2\pi)$.
On the closed orbit $\hat{n}$ is denoted by $\hat{n}_0$. 
In the rings treated here, $\hat{n}_0$ is nominally vertical in the arcs.
The rate of 
precession of spins around $\hat{n}$ is characterized
by the amplitude dependent spin tune $\nu_s$ \cite{D.P.Barber2004}.
This reduces to the closed-orbit spin tune, $\nu_0$, on the closed orbit. 
In a perfectly aligned planar ring, $\nu_0=a\gamma_0$, where
$a=0.00115965219$ for electrons(positrons), and $\gamma_0$ is the
relativistic factor for the design energy.
At orbital tunes for which constituent terms in the perturbation $\vec{\omega}$ stay coherent 
over long periods with the basic spin precession,  
$\hat{n}(\vec{u};\theta)$ can deviate strongly from $\hat{n}_0(\theta)$. This phenomenon is
called 
spin-orbit resonance (which will simply call ''spin resonance'' or ``resonance'' here) 
and the condition is 

\begin{equation}
    \label{spin resonances}
    \nu_s=k+k_x\nu_x+k_y\nu_y+k_z\nu_z, ~ ~ ~ k,~k_x,~k_y,~k_z\in\mathbb{Z} \; .
\end{equation}
where $\nu_x$, $\nu_y$ and $\nu_z$ are the orbital tunes. 
We expect that if, for some reason, a particle suffers large uncorrelated jumps
in its orbital phases, the coherence 
is lost and the resonances can be suppressed.

In storage rings, electron and positron beams become spontaneously
polarized due to the spin-flip synchrotron radiation, this phenomenon
is called the Sokolov-Ternov effect \cite{Sokolov:1963zn}. On the
other hand, the stochastic (synchrotron-radiation) photon emissions 
modify the trajectories and thus the $\vec{\omega}$
 along the trajectories. Then spin diffusion and thus depolarization
can occur \cite{V.N.Baier1966}. 
The polarization is therefore
a balance between the Sokolov-Ternov effect and the radiative
depolarization effect. In the seminal paper of Derbenev and
Kondratenko \cite{Derbenev:1973ia}, they derived the renowned formula
of the equilibrium beam polarization in an electron~(positron) storage
ring, which is along the direction of $\hat{n}$ at each point in phase space, and
with a magnitude of
\begin{eqnarray}
    \label{DK0}
    P_{\text{dk}}&=&-\frac{8}{5\sqrt{3}}\frac{\alpha_{-}}{\alpha_{+}} \nonumber \\
    \alpha_{-}&=&\oint{\mathrm{d}\theta}\langle\frac{1}{\lvert\rho{\rvert}^3}\hat{b}\cdot(\hat{n}-\frac{\partial\hat{n}}{\partial\delta})\rangle \nonumber \\
    \alpha_{+}&=&\oint{\mathrm{d}\theta}\langle\frac{1}{\lvert\rho{\rvert}^3}[1-\frac{2}{9}(\hat{n}\cdot\hat{\beta})^2+\frac{11}{18}(\frac{\partial\hat{n}}{\partial\delta})^2]\rangle,
\end{eqnarray}
where ${\partial\hat{n}}/{\partial\delta}$ is the so-called spin-orbit
coupling function, which quantifies the depolarization, and which can
be very large near spin resonances. $\hat{\beta}$ is a unit vector
along the direction of particle motion, $\dot{\hat{\beta}}$,
  the slope of $\hat{\beta}$ is therefore along the
  direction of acceleration, and 
  $\hat{b}=\hat{\beta}\times\dot{\hat{\beta}}/\lvert\dot{\hat{\beta}}\rvert$.
  The brackets ${\langle}{\rangle}$ denote an average over phase space
  at azimuth $\theta$. The term with ${\partial \hat n/ \partial \delta}$ in $\alpha_{-}$ accounts for the so-called
  kinetic polarization effect. In rings where ${\hat n}_0$ is nominally vertical in  the arcs
  this term is negligible. Following the definitions used in Ref. \cite{Derbenev:1979tm},
the polarization build-up rate is
\begin{equation}
    \lambda_{\text{dk}}=\lambda_{p}+\lambda_d^0,
\end{equation}
where $\lambda_p$ and $\lambda^0_d$ are the rates of the Sokolov-Ternov effect and the depolarization effect, respectively,
\begin{eqnarray}
    \label{DK0rate}
    \lambda_{p}&=&\frac{5\sqrt{3}}{8}\frac{r_e\gamma^5\hbar}{m_e}\frac{C}{2{\pi}c}{\oint}\mathrm{d}{\theta}\langle\frac{1-\frac{2}{9}(\hat{n}\cdot\hat{\beta})^2}{{\lvert}\rho{\rvert}^3}\rangle \nonumber \\
    \lambda^0_d&=&\frac{5\sqrt{3}}{8}\frac{r_e\gamma^5\hbar}{m_e}\frac{C}{2{\pi}c}{\oint}\mathrm{d}{\theta}\langle\frac{\frac{11}{18}(\frac{\partial\hat{n}}{\partial\delta})^2}{{\lvert}\rho{\rvert}^3}\rangle,
\end{eqnarray}
where $r_e$ and $m_e$ are the classical radius and mass of electron,
$\hbar$ is the Planck constant, $c$ is the speed of light, and $C$ is
the circumference of the storage ring. Note that $\lambda_p$ and $\lambda^0_d$ are 
dimensionaless, but that one can also define the 
characteristic time of the Sokolov-Ternov effect $\tau_p=C/(2\pi{c}\lambda_p)$ and of the
depolarization effect $\tau^0_d=C/(2\pi{c}\lambda^0_d)$, which can be compared with
the beam lifetime and aid the analysis of beam polarization. 
For later use we also need 
\begin{eqnarray}
{\tilde \lambda}_{p}&=&\frac{5\sqrt{3}}{8}\frac{r_e\gamma^5\hbar}{m_e}\frac{C}{2{\pi}c}{\oint}\mathrm{d}{\theta}\frac{1}{{\lvert}\rho{\rvert}^3}, \nonumber
\end{eqnarray}
but since the term with $\frac{2}{9}(\hat{n}\cdot\hat{\beta})^2$ 
makes a  negligible contribution for the rings considered here, ${\tilde \lambda}_{p} \approx \lambda_{p}$. 

A prerequisite for estimating the attainable beam
  polarization in an electron~(positron) storage ring is a detailed
  knowledge of the contributions to the depolarization. For that a
  simulation code must be established
  and that code can then serve as a guide for the design and
  optimization of the ring. Two classes of methods \cite{Chao:1999qt}
  have been developed for evaluating the equilibrium polarization in
  electron storage rings. 

One class computes $\hat{n}$ and ${\partial\hat{n}}/{\partial\delta}$
around the ring and applies Eq.~\ref{DK0} to obtain the equilibrium
beam polarization. Various codes of this class are listed in 
\cite{Chao:1999qt}, and they differ in the degree of linearization of
the spin and orbit motion, as well as being perturbative or
non-perturbative. The codes that handle linearized spin and orbit
motion, for example SLIM \cite{Chao:1980fz}, only describe 
the first-order spin resonances, namely those with $\nu_0 =
  k + k_x\nu_x + k_y\nu_y + k_z\nu_z$ for which,
  ${\lvert}k_x{\rvert}+{\lvert}k_y{\rvert}+{\lvert}k_z{\rvert}=1$,
  while higher-order spin resonances, namely those with
  ${\lvert}k_x{\rvert}+{\lvert}k_y{\rvert}+{\lvert}k_z{\rvert}>1$ are
  not taken into account. 
  Note that higher-order spin resonances stem from the
    three dimensional nature of spin motion and that they can be
    driven by linear orbital motion. This is ultimately 
    due to the fact that successive spin rotations around different axes do not commute. This is illustrated 
    in the SMILE formalism \cite{Mane:1986qz} which evaluates the equilibrium
    polarization of Eq.~\ref{DK0} using a 
    perturbative calculation of $\hat{n}$ and
    ${\partial\hat{n}}/{\partial\delta}$ in power series of the
    orbital amplitudes. The SMILE
    formalism also illustrates how to identify the sources of
    resonances and, importantly,  how to classify them. Note also that for
    perturbative calculations it is $\nu_0$ that appears in the
    expression for a resonance, not $\nu_s$. 
  The normal form analysis in PTC can also be used to compute
  $\hat{n}$ on particle trajectories perturbatively \cite{icap2012}.
  However, these perturbative approaches have convergence problems for
  the calculation of synchrotron sideband resonances for high
  beam energies \cite{S.R.ManePrivate}. 

   Synchrotron sideband resonances are
    those resulting from the modulation of the rate of spin precession
    around ${\hat n}_0$ due to the energy oscillations inherent in
    synchrotron oscillations. For example, for first-order parent
    $\nu_x$ or $\nu_y$ resonances the sideband resonances satisfy the
    condition $\nu_0 = k + k_x\nu_x + k_y\nu_y + k_z\nu_z$ for which,
    ${\lvert}k_x{\rvert}+{\lvert}k_y{\rvert}=1$ with $k_z \ne 0$. The
    occurrence of extra resonances is easy to understand when it is
    recalled that the modulation of a base frequency always introduces
    sidebands into the frequency spectrum. The modulation of the rate
    of spin precession around ${\hat n}_0$ due to energy oscillations
    mostly occurs in the arcs where ${\hat n}_0$ is nominally vertical
    and where particles experience a contribution to their horizontal
    motion in the quadrupoles due horizontal dispersion and energy
    oscillations. A modulation also occurs due to horizontal betatron
    motion in arc quadrupoles. However, betatron tunes are large so
    that the average modulation over one turn is small and then the
    corresponding sidebands are insignificant. In contrast,
    synchrotron tunes are very small so that the average modulation
    cannot be ignored. In any case we focus here on synchrotron motion because 
    it is this motion that is in first instance affected by photon emission.

SODOM
 \cite{Yokoya:1992nt} treats linearized orbit motion and
three-dimensional spin motion. The field $\hat{n}$ is computed
non-perturbatively and a first-order difference of $\hat{n}$ at
several nearby phase space points is used to compute
${\partial\hat{n}}/{\partial\delta}$. In principle, other
non-perturbative algorithms for calculating $\hat{n}$ can be used in a
similar way, for example the stroboscopic averaging method \cite{K.Heinemann1996}.
These non-perturbative algorithms can treat the synchrotron sideband resonances correctly.

The other class simulates the depolarization due to non-spin-flip
synchrotron radiation using a Monte-Carlo technique to compute the
depolarization rate $\lambda_d$, and it is a good approximation to 
ignore the tiny effect of kinetic polarization and estimate the equilibrium polarization in an electron~(positron) storage
ring following
\begin{eqnarray}
    \label{equilibrium polarization II}
    P_{\text{eq}}&\approx&\frac{P_{\infty}}{1+\lambda_d/\lambda_p} \nonumber \\
    P_{\infty}&\approx&-\frac{8}{5\sqrt{3}}\frac{\oint{\mathrm{d}\theta} \frac{1}{\lvert\rho{\rvert}^3}\hat{b}\cdot\hat{n}_0}{\oint{\mathrm{d}\theta} \frac{1}{\lvert\rho{\rvert}^3} [1-\frac{2}{9}({\hat{n}}_0\cdot\hat {\beta} )^2 ]  } 
\end{eqnarray}
where $P_{\infty}$ is the equilibrium polarization taking into account
the orbital imperfections, but disregarding the depolarization effects
due to stochastic emission of synchrotron radiation. This term can be
computed using a linear code like SLIM \cite{Chao:1980fz}.

In this approach, the first-order and higher-order spin resonances are
automatically handled since it is based on three-dimensional spin
motion. The Monte-Carlo method does not rely on
  calculating $\hat n$ and its derivative and in any case it is valid
  independently of whether Eq.~\ref{DK0} or some other description of
  spin diffusion is valid. It can thus be used to check the
  theoretical models of the equilibrium beam polarization. 

SITROS was the first code to use this pragmatic approach. It was first
developed by Kewisch \cite{Kewisch:1983uq,Kewisch:1988nv} and later
upgraded by Boge \cite{M.Boge} and Berglund 
\cite{M.Berglund}. It divided a storage ring into sections,
and the break points were bending magnets where generation of ''big
photons'' was simulated, and interaction points where weak{\textendash}strong 
beam{\textendash}beam interactions were included. Particles were transported 
using second-order transfer maps which included chromatic effects and the effects of 
sextupoles but which were not exactly symplectic.
SITROS was used for HERA and LEP. 
Barber's SLICKTRACK
  \cite{Barber:2005slicktrack} code also implements the ''big
  photons'' between sections, but transports particles between these
  break points with the thick-lens symplectic transfer matrices of
  SLIM. Implementation of the nonlinear orbital motion is one of its
upgrade plans. Currently SLICKTRACK is well maintained and widely
used, for example in the design study of electron{\textendash}ion colliders and
ILC damping rings.

With the great development of computing power since SITROS was first
developed, it is now possible to implement such a Monte-Carlo
simulation in a more ''natural'' way, namely with distributed
generation of synchrotron-radiation photons in each integration step.
Moreover, nonlinear orbital motion can be taken into account with the
implementation of symplectic integrators in modern tracking codes.
PTC is a tracking code developed by E. Forest \cite{Schmidt:2002vp},
which was designed to model various geometries of particle
accelerators, and do symplectic tracking of the orbital motion and
length-preserving transport of spin \cite{S.Mane2009PTC}. Particle
coordinates and Taylor maps can be tracked in a polymorphic manner,
and the latter enables the normal form analysis of the one-turn map
using FPP \cite{fppipac2006}. PTC is now embedded in MADX \cite{MADX}
and BMAD \cite{Sagan:2006sy}, and some of its functionalities can be
called in MADX and BMAD as a library. On the other hand, complicated
operations like implementation and correction of machine imperfections
can be done using MADX or BMAD, and the lattice is then dumped to an
input file for PTC. Tracking studies can be carried out inside PTC
afterwards. Therefore, PTC has an ideal framework for the
implementation of a Monte-Carlo simulation of the equilibrium beam
polarization in electron storage rings.

The implementation of a Monte-Carlo simulation in PTC is described in
section \ref{sectionI} together with its benchmarking against SODOM
\cite{Yokoya:1992nt}. In section \ref{sectionII}, we review the
theory of beam polarization for ultra-high beam energies. In section
\ref{sectionIII}, a case study of the beam polarization at ultra-high
beam energies is presented using this Monte-Carlo approach for a model
storage ring, and compared with the theory.

\section{\label{sectionI} Code implementation and benchmarking}
\begin{figure}[htb!]
        \begin{center}
            \includegraphics{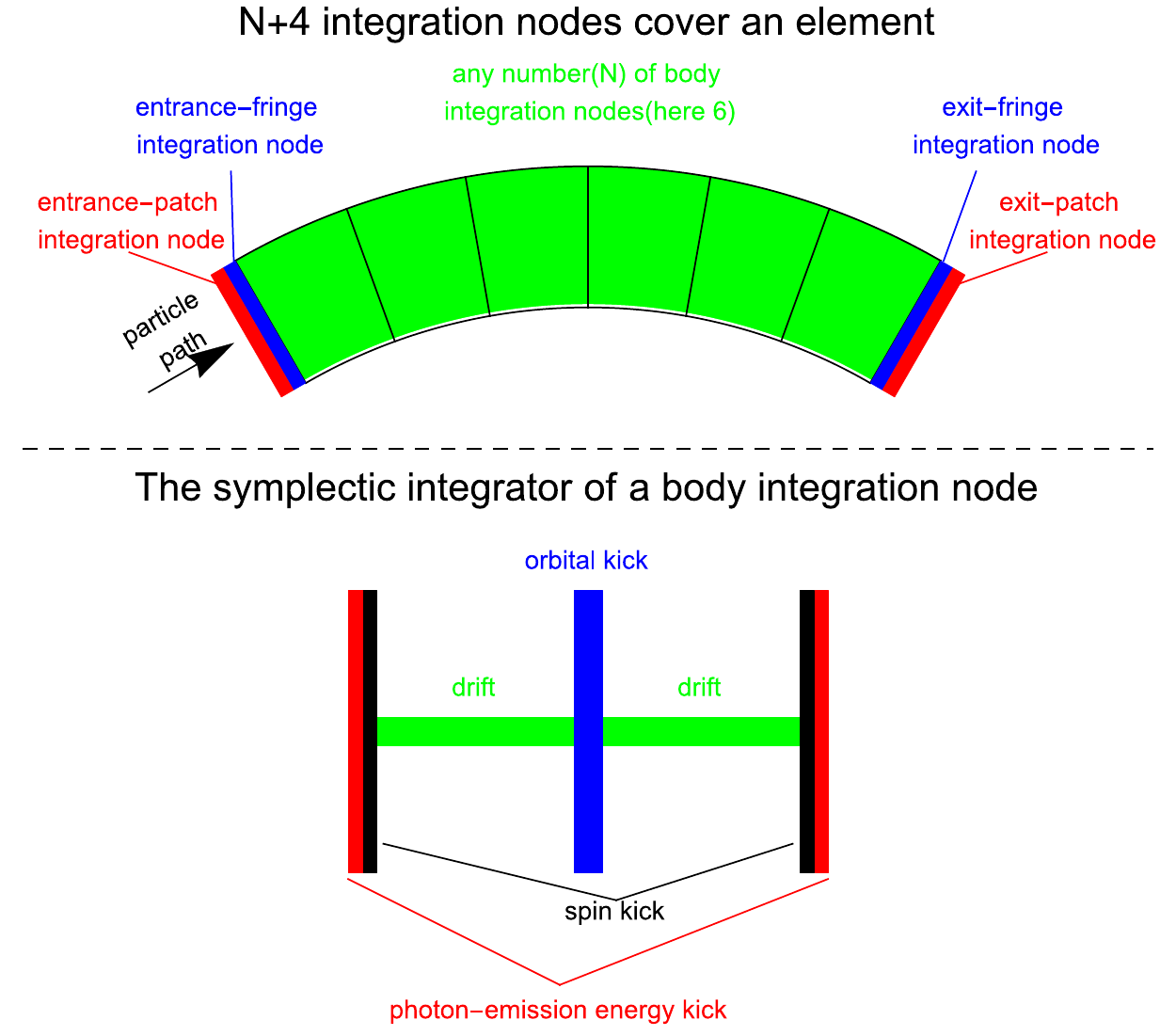}
        \end{center}
        \caption{A schematic diagram  showing the tracking scheme in PTC. The top part shows the
        integration nodes inside an element \cite{PTClibug}, while the bottom part shows the scheme of symplectic 
        integration in a body integration node, and photon emission.}
        \label{fig:PTCschematic}
    \end{figure}

When a particle (or a Taylor map) is tracked through an element in PTC, a step
of integration is called  an ``integration node''. There are five different
types of integration nodes, describing the entrance patch, the entrance fringe
field, the body of an element, the exit fringe field and the exit patch. Here,
patches connect the local coordinates of adjacent elements. The body of an
element can be split into a number of integration nodes. A schematic diagram of
the integration nodes inside an element is shown in Figure
\ref{fig:PTCschematic}.
    
In this context, synchrotron radiation effects are only taken into
account for the integration nodes in the body of bending magnets. When a
particle is tracked through an integration node of a magnet body, the miniscule
Stern{\textendash}Gerlach effect is neglected and 
the orbital transfer map is sandwiched in between two spin kicks in equal
amounts, and they together form a second-order symplectic
integrator for the particle motion and a length-preserving rotation for the spins. 
The kick due to photon emission is
placed next to the spin kick, and they commute with each other, as
the spin-dependent synchrotron radiation is not taken into
account. Therefore the synchrotron-radiation process is also
concentrated before the entrance and after the exit of the orbital
map in equal amounts. 
The transport of orbital coordinates in each integration node can be chosen
from the second-order, the fourth-order and the sixth-order symplectic
integrators.  In this simulation study, we choose the second-order symplectic
integrator. Because the quadrupoles are split into many integration nodes to
ensure the accuracy of spin tracking, and dipoles are also split so that the
number of photons emitted in each integration step is small, and the desired
accuracy of orbital motion can be achieved without going to a higher-order
symplectic integrator.  The tracking scheme for an integration node of the body
of a magnet is shown in Figure \ref{fig:PTCschematic}. 

In each integration step, the local radius of curvature is
computed for each tracked particle, as well as the critical energy
$u_c$. 
The number of emitted photons
$n_\gamma$ is first randomly generated with a Poisson distribution.
Let the energy of an emitted photon be $u_0$, then the relative energy $\xi=u/u_c$ 
is stochastically generated following the algorithm implemented in GEANT4
\cite{Burkhardt:2007zza}, which is quite fast and more precise than
former implementations \cite{Roy:1990vw}. The stochastic energy kick
before or after the orbital kick is
\begin{equation}
    \label{quantum excitation}
    \delta\rightarrow\delta-\sum_{1}^{n_\gamma}{\xi}u_c/E_{beam}.
\end{equation}

A bunch of particles are launched on the closed orbit with spins
initialized along the $\hat{n}_0$, and are then tracked for several
damping time. The beam polarization $P(t)$ as the ensemble average of
these particle spins, is computed during the tracking. The
depolarization rate is fitted following
\begin{equation}
    \label{depolarization rate}
    P(t)=\exp(-\frac{2{\pi}ct}{C}{\lambda}_d),
\end{equation}
and Eq.~\ref{equilibrium polarization II} is then used to compute the equilibrium beam polarization.

The calculations with this method are benchmarked against SODOM
\cite{Yokoya:1992nt} for the same model storage ring (Model 1) as
described in Ref. \cite{Yokoya:1992nt}. Model 1 consists of 128
identical FODO cells and four FODO cells with vertical bends (four
upward bends and four downward bends with a bending angle of 0.017453
rad, to introduce vertical dispersion). All the dipoles are 6 m
long, the quadrupoles are 1 m long with inverse focal lengths of
$k_f=0.18243216~\text{m}^{-1}$ and $k_d=0.16763610~\text{m}^{-1}$,
respectively, and the cell length is 16 m. A half ring consists
of an upward-bend cell, 32 FODO cells, an RF cavity~(zero length), 32
FODO cells and an upward-bend cell. In the other half ring, the
upward-bend cells are replaced by downward-bend cells. The parameters
are listed in Table \ref{SODOM_parameters}. No machine imperfections
are introduced in this study.
\begin{table}[htb]\footnotesize
\centering
\caption{\label{SODOM_parameters} The parameters of Model 1 storage ring}

\begin{tabular}{cc}
        \toprule
\textbf{Parameter} &  \textbf{Value} \\
        \midrule
        Circumference(m) & 2112    \\
        Beam energy(GeV) & 20.736  \\
        $\nu_x/\nu_y/\nu_z$ & $33.265/28.380/0.0623$ \\
        Relative energy spread   & $1.13\times10^{-3}$ \\
        Damping time(turns) & 620/620/310  \\
        Emittance($\text{mm}{\cdot}\text{mrad}$) & $6.03\times10^{-2}/2.38\times10^{-4}/7.42$ \\
        Sokolov-Ternov time($\tau_p$)(minute) &  8.6 \\
        Rms spin precession frequency spread ($\sigma_{\nu}$) & 0.053    \\
        Modulation index($\alpha$) & 0.732     \\
        Correlation index($\kappa$) &   0.012  \\
        \bottomrule
\end{tabular}
\end{table}

\begin{figure}[p]
\begin{subfigure}{\linewidth}
\centering
\includegraphics[width=80mm]{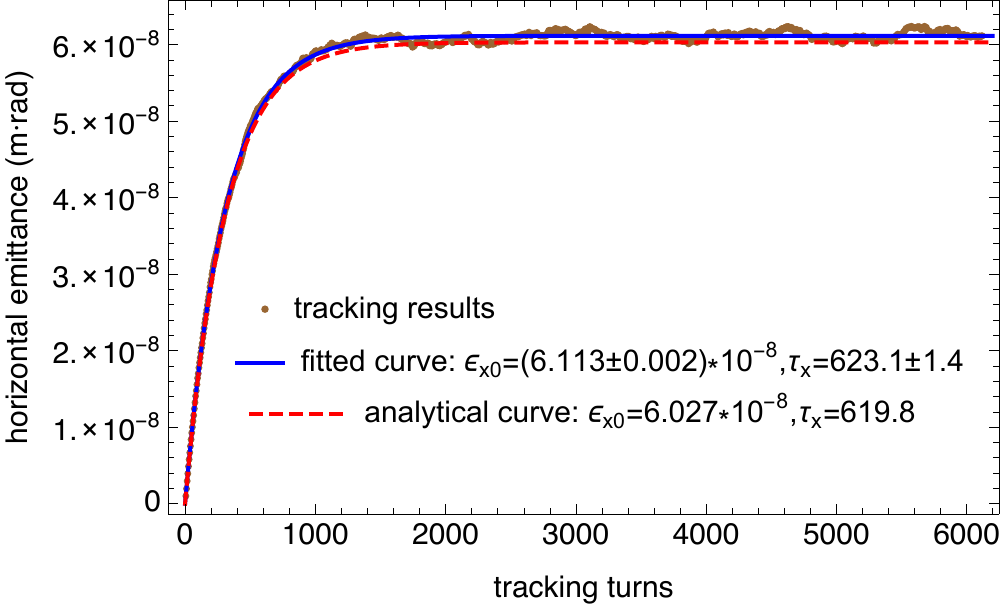}
\caption{horizontal emittance}
\label{fig:emitx}
\end{subfigure}\\[1ex]
\begin{subfigure}{\linewidth}
\centering
\includegraphics[width=80mm]{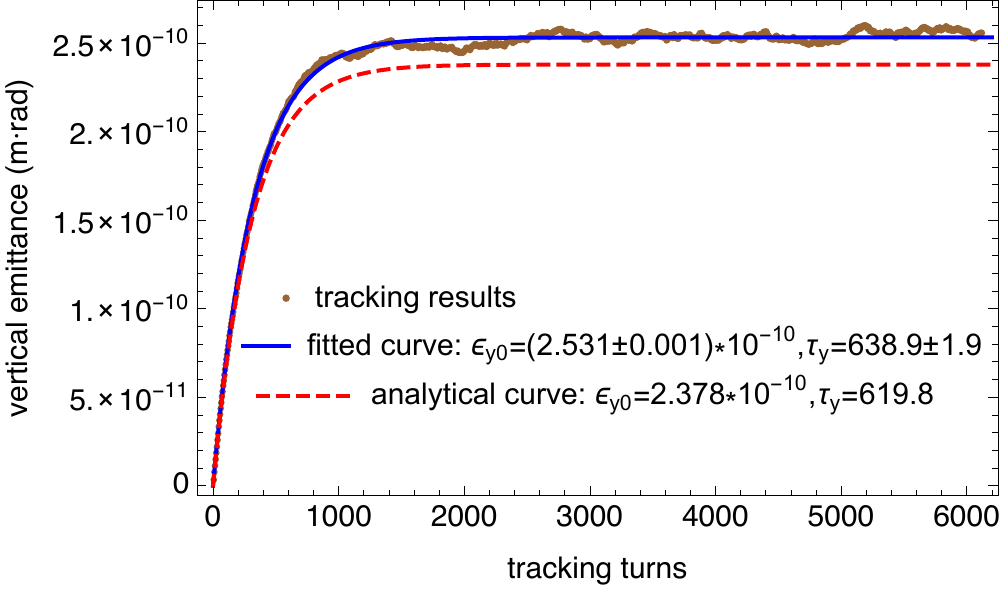}
\caption{vertical emittance}
\label{fig:emity}
\end{subfigure}\\[1ex]
\begin{subfigure}{\linewidth}
\centering
\includegraphics[width=80mm]{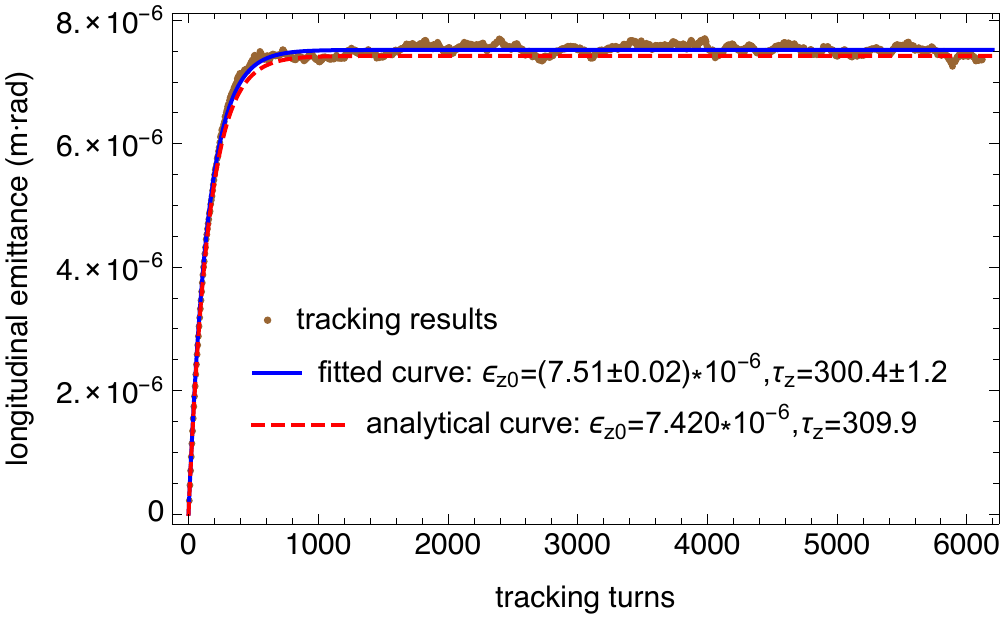}
\caption{longitudinal emittance}
\label{fig:emitz}
\end{subfigure}
\caption{
Evolution of eigen-emittances of a beam of 9600 particles in the Monte-Carlo simulation for Model 1. 
The evolutions of the emittances are fitted following $\epsilon_i(t)=\epsilon_{i0}(1-\exp(-t/\tau_i)),\ i=x,y,z$. Comparison of the fitted parameters and the analytical solutions are also shown in the figures.}
\label{SODOM emittance}
\end{figure}

\begin{figure}[htb!]
\centering
\includegraphics[width=130mm]{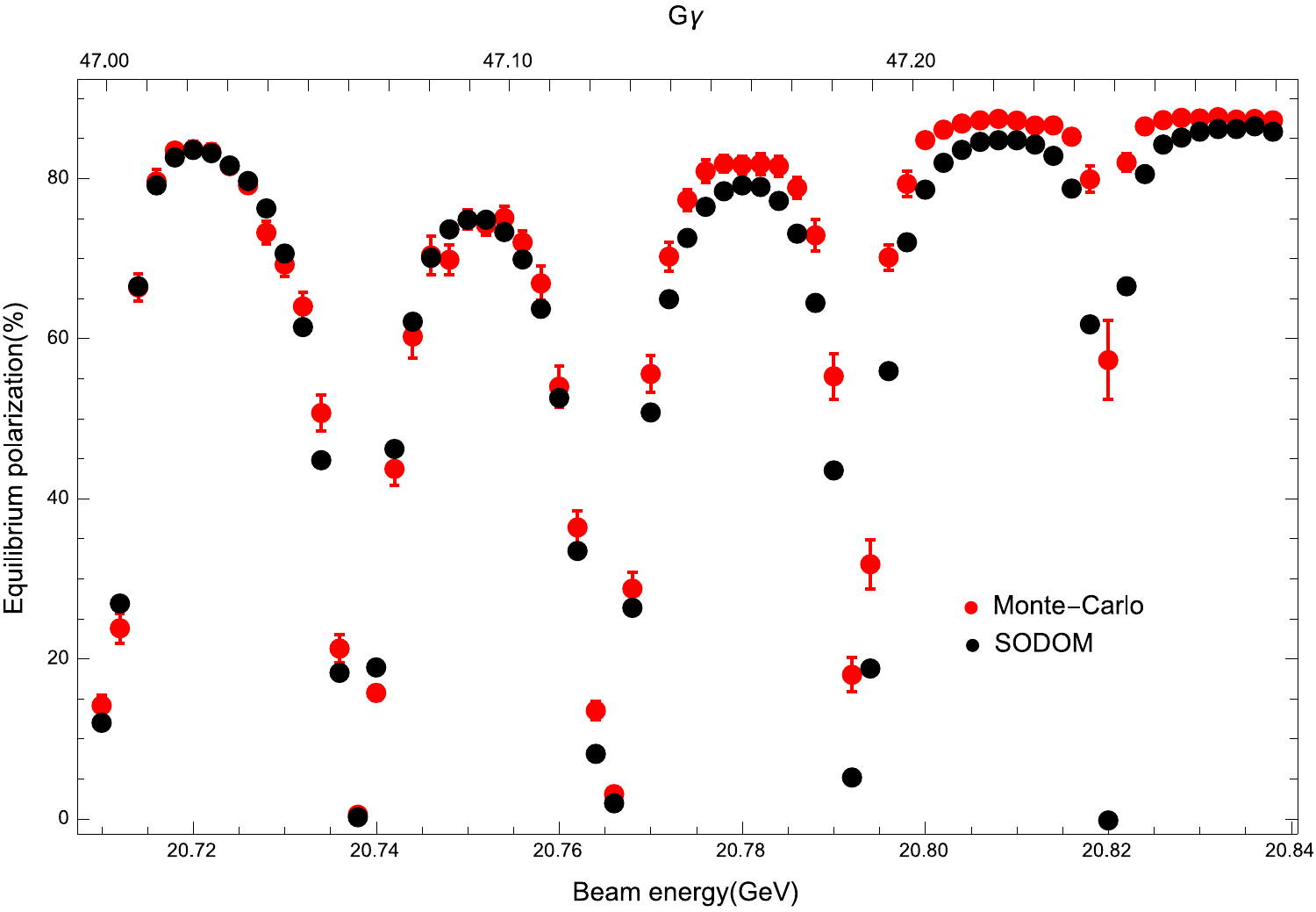}
\caption{Comparison of the computed equilibrium polarizations for Model 1. ''SODOM'' is taken from the Yokoya's paper \cite{Yokoya:1992nt} with his permission. ''Monte-Carlo'' is the Monte-Carlo simulation result with 50 particles, the statistical error is calculated with 20 such simulations. The agreement is good.} 
\label{SODOM benchmark}
\end{figure}

The first step is to check the evolution of the beam's
eigen-emittances \cite{Alexahin:2014cha} computed using the tracking
data of a beam of 9600 particles. These are then compared with the
analytical solutions, where the beam envelope formalism \cite{ohmienv}
is used to compute the equilibrium beam emittances. As 
shown in Figure \ref{SODOM emittance}, the fitted equilibrium 
emittances and damping times differ from the analytical results by 
just a few percent.
The equilibrium polarization is calculated for the energy range of
20.71\textendash20.84~GeV, where several synchrotron-sideband spin
resonances are visible. The expectation number of the emitted photons is
$\sim11$ for the main dipole. Therefore, the number of integration
steps is chosen to be 10 and it is unlikely that a large number of
photons are generated at each integration node. It takes around a
minute to track a particle for 3000 turns (5 damping times) using CPUs
of the Hopper cluster at the National Energy Research Scientific
Computing Center~(NERSC), and the tracking of different particles can
be parallelized. Several simulations with the same number of
particles can be launched. The depolarization rate is the average of
the results of these simulations, and the statistical error can also
be calculated. A comparison of the Monte-Carlo result with 50
particles and SODOM is shown in Figure \ref{SODOM benchmark}. The
results are consistent with each other.

There are several possible issues for this Monte-Carlo simulation.
First, if the depolarization rate is very low then it would take an
impractically long time to reach an accurate estimation of the
depolarization rate and hence the equilibrium beam polarization.
Second, since the
particles are launched on the closed orbit, it takes many turns for
particles to diffuse to large amplitudes, and the contribution from
phase space with large amplitudes is likely to be underestimated. To
solve this problem, the beam can be given a Gaussian distribution at
the start \cite{Kewisch:1988nv}. Note that the particle spins should
then be initialized to be parallel to the ISF. In addition, it is
arguable whether implementation of nonlinear orbit motion and photon
emission at each integration step are necessary for all practical
simulations. However, it is not trivial to turn off the nonlinear
orbit motion in the tracking. Then the linear transfer matrices
obtained by tracking the first-order Taylor map could be used instead
and the algorithm of SLICKTRACK then realized with some effort.
Moreover, the number of integration steps of bending magnets can be
easily set in PTC, and computing time can be reduced when needed.
    
\section{Further details of the theory  \label{sectionII}}
    Electron~(positron) polarization has been observed and studied in
several storage rings \cite{S.R.Mane2005, Barber:1994ew}  
and the polarizations attained have been broadly in line with the expectations of Eq.~\ref{DK0}.
In storage rings with relatively low beam energies, the first-order
spin resonances of Eq.~\ref{spin resonances} dominate together with their synchrotron sidebands. 
Resonances can have several origins including transverse coupling, lack of so-called 
spin transparency in regions where ${\hat n}_0$ is horizontal due to the use of spin rotators,       
and indirect effects resulting from orbital imperfections. Various spin matching
schemes \cite{Chao:1999qt} have been studied and implemented in
several machines to weaken the first-order resonances and these schemes bring improvements in the achievable
equilibrium polarization. In high energy storage rings, it is
generally believed that the synchrotron sideband resonances are
the most important family of spin resonances.

We have already described how synchrotron sideband resonances have their origin 
in the modulation of the spin precession rate due to synchrotron motion.
We now look in more detail.

As mentioned earlier, near spin-orbit resonance,  $\hat{n}$ can be strongly spread
out away from ${\hat{n}}_0$. Then, even in the absence of
radiation, the maximum attainable equilibrium polarization of the
beam, $\langle \hat n \rangle$, can be be small. In particular, in
the case of synchrotron sideband resonances, $\hat{n}$ can vary
strongly as a function of the synchrotron phase with the same
orbital actions and this can contribute to making the spin-orbit
coupling function large. Then the radiative depolarization can be
strong and lead to small equilibrium beam polarization according
to Eq.~\ref{DK0}. Here, we have used a standard picture in which we
treat the betatron and synchrotron motion on an equal footing.
However, we can gain extra insight by treating the betatron and
synchrotron motion separately and by noting that if the modulation
of the spin precession rate causes an initially vertical spin to
cross a parent betatron spin resonance, the spin can be strongly
disturbed and might even flip over as expected from the
Froissart-Stora formula \cite{Froissart1960} and as illustrated in \cite[pp.74-75]{Hoff's book}. 
Although this argument is heuristic in that it 
likens the modulation of the spin precession frequency to a modulation 
of the closed orbit spin tune $\nu_0$, it contains the essentials. Note that in this description 
the typical rate of resonance crossing  increases with $\nu_z$.
Both of these pictures assume long-term coherence or near coherence between relevant terms in
$\vec{\omega}$ and the basic spin motion. 

Note that the
introduction of Siberian snakes \cite{Derbenev1977,
  Derbenev1978} might reduce the dependence of spin
transport on the energy oscillations of synchrotron motion so that
the mechanism could be suppressed \cite{Barberspin2010}. This
topic is beyond the scope of this paper.

The synchrotron sideband resonances of an integer resonance are
$\nu_0=k+m\nu_z$, with the contribution to
${\lambda^0_d}/{{\tilde \lambda}_p}$ \cite{Mane:1990ge}
\begin{equation}
    \label{int sideband}
    \frac{\lambda^0_d}{\tilde{\lambda}_p}=A\sum\limits^{\infty}_{m=-\infty}\left[\frac{\Delta\nu}{(\Delta\nu+m\nu_z)^2-\nu_z^2}\right]^2e^{-\alpha}I_m({\alpha}),
\end{equation}
where $\Delta\nu$ is the distance between the closed orbit spin
tune and the parent resonance, say $\Delta\nu=\nu_0-k$, $\lvert{m}\rvert$ is the order of the sideband
resonance, $I_m$ is a modified Bessel function, $\sigma_{\epsilon}$ is the rms relative energy spread, and 
$\alpha=(\nu_0\sigma_{\epsilon}/\nu_z)^2$ the tune modulation
index. $A$ is a constant that relates to the width of the first-order
synchrotron sideband resonance, because if we set
$\alpha=0$, then
\begin{equation}
   \label{linear sync sideband}
    \left.\frac{\lambda^0_d}{\tilde{\lambda}_p}\right|_{\alpha=0}=\frac{A\Delta\nu^2}{(\Delta\nu^2-\nu^2_z)^2}.
\end{equation}

The synchrotron sideband resonances of an isolated horizontal
parent resonance are $\nu_0=k\pm\nu_x+m\nu_z$, with the
contribution to ${\lambda^0_d}/{\tilde{\lambda}_p}$ \cite{Mane:1990ge}
\begin{align}
    \label{beta sideband}
    \frac{\lambda^0_d}{{\tilde \lambda}_p}=B\sum\limits^{\infty}_{m=-\infty}&\left\{\frac{e^{-\alpha}}{(\Delta\nu+m\nu_z)^2}\times\right. \nonumber\\
    &\left[J_xI_m(\alpha)+J_z\frac{\alpha}{2}(I_{m-1}(\alpha)+I_{m+1}(\alpha))\right]+\nonumber\\
    &\left.\frac{e^{-\alpha}}{(\Delta\nu+m\nu_z)}\frac{mJ_z}{\nu_z}I_m(\alpha)\right\},
\end{align}
where $\Delta\nu=\nu_0-(k\pm\nu_x)$, $J_x$ and $J_z$ are the horizontal and longitudinal damping
partition numbers and $B$ is a constant that relates to the width of the
horizontal parent resonance. If we set $\alpha=0$, then
    \begin{equation}
        \label{linear bet sideband}
        \left.\frac{\lambda^0_d}{\tilde{\lambda}_p}\right|_{\alpha=0}=\frac{BJ_x}{\Delta\nu^2}.
    \end{equation}
The result for synchrotron sidebands for an isolated vertical
parent resonance can be obtained by replacing $x$ by $y$ in Eq.
12. The equilibrium polarization can then be estimated as
\begin{equation}
    \label{DK1}
    P_{\text{eq}}\approx~\frac{P_0}{1+\sum\limits_{\nu_k}\left(\frac{\lambda^0_d}{\tilde{\lambda}_p}\right)_{\nu_k}},
\end{equation}
where $\nu_k=k_0+k_x\nu_x+k_y\nu_y+k_z\nu_z$ is the location of a parent
resonance. Therefore, once we have computed the equilibrium
polarization using a linear formalism like SLIM, the widths $A$ and
$B$ of the first-order (betatron and synchrotron) spin resonances can be fitted and the
equilibrium polarization taking into account the synchrotron
sideband resonances can then be estimated analytically using Eq.~\ref{DK1}.
Note, however, that the nearby first-order horizontal and
vertical parent spin resonances might not be well separated and that resulting interference effects are not
included in Eq.~\ref{DK1}.

 Note that these calculations of the strengths of the
synchrotron sideband resonances are based on evaluation of the
spin{\textendash}orbit coupling function in Eq.~\ref{DK0} and that assumes that
the synchrotron motion is well defined so that the synchrotron
phases are strongly correlated from turn to turn.
However, it has been suggested that for ultra-high beam energies, it is possible to enter a
new regime whereby the successive passages of spin resonances during  synchrotron oscillations are
uncorrelated \cite{Derbenev:1979tm} and that this can modify the rate of depolarization. 

In fact in Ref. \cite{Derbenev1975}, it was argued that Eq.~\ref{DK0} should be
generalized when the system is very 
close to a spin resonance. In particular, following the  heuristic physical explanation
in Ref. \cite{Kondratenko1974}, very close to a spin resonance,
stochastic photon emissions should contribute a depolarization rate of  
\begin{equation}
\label{deprate}
\lambda'_d=\pi\sum\limits_k\langle\lvert\omega_k\rvert^2\delta(\nu_s-\nu_k)\rangle.
\end{equation}
So an additional term was added to $\alpha_+$
in Eq.~\ref{DK0} when evaluating the equilibrium beam polarization to give
\begin{align}
    \label{newDK0}
    \alpha_{+}=&\oint{\mathrm{d}\theta}\langle\frac{1}{\lvert\rho{\rvert}^3}[1-\frac{2}{9}(\hat{n}\cdot\hat{\beta})^2+\frac{11}{18}(\frac{\partial\hat{n}}{\partial\delta})^2]\rangle+ \nonumber \\
    &\frac{8}{5\sqrt{3}}\frac{2{\pi}c}{C}\frac{m_e}{r_e{\gamma}^5\hbar}\pi\sum\limits_k\langle\lvert\omega_k\rvert^2\delta(\nu-\nu_k)\rangle.
\end{align}
where the perturbed spin precessing frequency is $\nu=\nu_0+\Delta\nu$ with $\Delta\nu$ due to the energy oscillation, $\nu_k=k+k_x\nu_x+k_y\nu_y+k_z\nu_z$ is the location of a spin
resonance, and $\omega_k$ is the amplitude of the Fourier component for the first-order
 spin resonances. The
average in the additional term is taken over the beam distribution. 

In summary, it was suggested  that the total depolarization rate should be
$\lambda_d=\lambda^0_d+\lambda'_d$. The additional term should  not be
important if the spread of the spin precession frequency is small,
because we do not run a machine very close to a major spin resonance. 

Now let us estimate the spread in the spin phases with synchrotron oscillations
and stochastic photon emission. Following the derivation in Ref.
\cite{Yokoya:1983sm}, the variation of the spin precession phase
$\Phi$ can be expressed as
     \begin{equation}
         \label{spin precession}
         \frac{d\Phi}{d\theta}=a\gamma=a\gamma_0(1+\delta),
     \end{equation}
 Let $\Delta\delta_i$ be the change of $\delta$ due to the emission of a 
 photon at $\theta=\theta_i$. Then the spin phase
 will be shifted by the amount  
 \begin{equation}
     \label{spin phase variation}
     a\gamma_0\Delta\delta_i\int\limits_{\theta_i}^{\theta}\cos\nu_z(\theta'-\theta_i)d\theta'=\frac{\nu_0}{\nu_z}\Delta\delta_i\sin\nu_z(\theta-\theta_i),
 \end{equation}
at the azimuthal angle $\theta$ after the emission. Then summing
up all the effects of photon emissions between $\theta'=0$ and
$\theta'=\theta$, the total variation of spin phase is
 \begin{equation}
     \label{total variation of phase}
     \Delta\Phi(\theta)=\frac{\nu_0}{\nu_z}\sum\limits_{0<\theta_i<\theta}\Delta\delta_i\sin\nu_z(\theta-\theta_i).
 \end{equation}
Averaging $(\Delta\Phi(\theta))^2$ 
on all the emissions that occur between $\theta'=0$ and
$\theta'=\theta$,
 \begin{equation}
     \label{averaging}
     \langle(\Delta\Phi(\theta))^2\rangle=(\frac{\nu_0}{\nu_z})^2\langle(\Delta\delta_i)^2\rangle\frac{dN}{d\theta}\int_0^{\theta}d\theta'\sin^2\nu_z(\theta-\theta').
 \end{equation}
 Here $dN/d{\theta}$ is the mean number of emitted photons during one radian of $\theta$.
 Since \cite{Sands:1970ye}
 \begin{equation}
     \label{integral}
     \frac{dN}{d\theta}\langle(\Delta\delta_i)^2\rangle=\frac{11}{9}{\tilde \lambda}_p,
 \end{equation}
 we then get the spread in spin phase in a synchrotron oscillation period
\begin{equation}
     \label{averaging1}
     \langle(\Delta\Phi)^2\rangle=\frac{11}{18}(\frac{\nu_0^2}{\nu_z^3}){\tilde \lambda}_p.
 \end{equation}
 We define the correlation index as $\kappa=\frac{11}{18}{\nu_0^2{\tilde \lambda}_p}/{\nu_z^3}$.
 If $\kappa\ll1$, the spread of spin phase is small, 
 then the successive
 passages of the spin resonance due to synchrotron oscillation are
 correlated.

 Otherwise, if the rms spread of the spin-precession frequency
 $\sigma_{\nu}=\nu_0\sigma_{\delta}{\gg}\nu_z$, the crossings of
 resonances during synchrotron motion are completely uncorrelated.
 It is claimed that in this regime the synchrotron oscillation plays the role of SR
 photon emission in driving the uncorrelated resonance crossings,
 and that the depolarization rate is described by Eq.~\ref{deprate}.
 Moreover, the spin resonances completely overlap and are
 unavoidable. Then it is not clear to us whether the analytical calculation  of Eq.~\ref{DK0} for
 $\lambda^0_d$ is still applicable and, if it is not applicable, it is not clear to us what prescription would be 
 used instead. Furthermore it is not clear to us whether
 the total depolarization rate can be obtained by simply adding
 $\lambda^0_d$ and $\lambda'_d$ together.

 In Ref. \cite{Derbenev:1979tm}, it was suggested that when
 $\sigma_{\nu}\ll1$, $\lambda'_d$ and $\lambda^0_d$ would be
 comparable, and should be added together to obtain the
 depolarization rate $\lambda_d$. Moreover, in Ref. 
 \cite{Derbenev:1979tm} only the contributions from the parent
 resonances were used to calculate $\lambda^0_d$.
 When $\sigma_{\nu}\gg1$, $\lambda^0_d$ is negligibly small and it
 was predicted there there would be no resonant dependence of spin
 diffusion on energy, and that beam polarization would increase with
 energy. 
 
 The above idea was applied to the LEP storage ring 
 \cite{Assmann} and it was shown that the depolarization should have entered
 the uncorrelated regime for LEP above 60~GeV. A new optics was
 implemented in LEP which facilitated establishing polarization at
 60.6~GeV \cite{101/45optics}, whereas a beam polarization of 
 below $1\%$ was measured at 70~GeV, 92~GeV and 98.5~GeV, in
 qualitative agreement with Eq.~\ref{DK0}. So the hypothetical
 increase of polarization at high energy was not observed.

\section{\label{sectionIII} Simulation study of a model storage ring}

\begin{figure}[htb!]
\centering
\includegraphics[width=130mm]{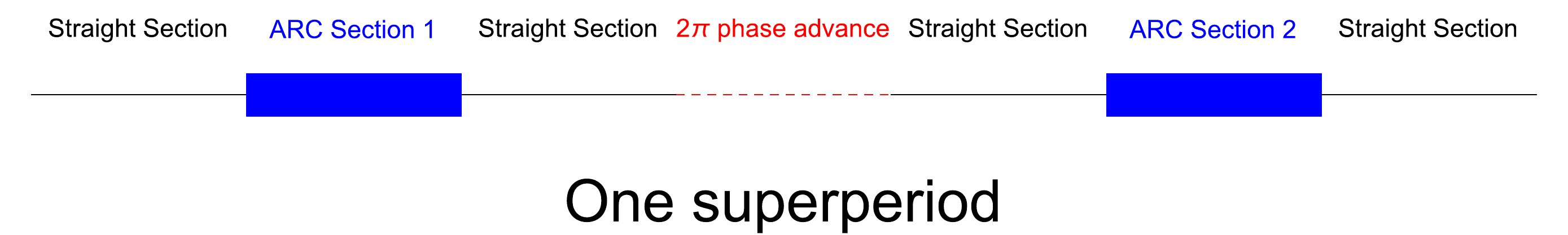}
\caption{A schematic plot of a superperiod of the Model-2 storage ring. ARC sections 1 and 2 are identical. Therefore
two identical straight{\textendash}arc{\textendash}straight sections are connected with a section with $2\pi$ phase advance.} 
\label{Superperiod}
\end{figure}

For this study we adopted a simplified version of the CEPC lattice
(Model 2). It has a periodicity of $P=4$, with one superperiod shown
in Figure \ref{Superperiod}. 
The arc sections are composed of FODO
cells with $60^{\circ}$ phase advance in both transverse planes and
they have dispersion suppressors with missing dipoles. The simplified straight
sections are composed of FODO cells without taking the
injection insertions into account. Identical RF cavities of zero length are placed
in the center of each straight section. The ''$2\pi$ phase advance''
section occupies the space left for the final focus insertion. Four
identical thin skew quadrupoles are inserted at the center of each
superperiod and the transverse emittance ratio is tuned to $0.003$.
No orbit imperfections are added in the simulations, and the first-order 
synchrotron sideband resonances are generally weak. 
For a four-fold symmetric ring the parent betatron spin resonances 
are in principal located at 
\begin{equation}
\nu_0=4k~{\pm}~\nu_j,~ ~ j=x,~y~ ~\text{and} ~k\in\mathbb{Z}.
\end{equation}

In this section, two cases of different energy ranges will be studied
around 120~GeV and 150~GeV, respectively. The parameter list relevant
for this study is shown in Table \ref{CEPC_parameters}. 
Note that for these parameters we take into account the effect of 
the sawtooth shape of the closed orbit caused by synchrotron radiation in the dispersive regions of the ring. 
\begin{table}[htb]\scriptsize
\caption{\label{CEPC_parameters} The parameters of Model 2 storage ring}
\begin{tabular}{ccc}
        \toprule
\textbf{Parameter} &  \textbf{Case 1} & \textbf{Case 2}\\
        \midrule
        Circumference(m) &  54752 & 54752 \\
        Beam energy(GeV) &  120 & 150 \\
        $\nu_x/\nu_y/\nu_z$ &  $193.084/193.218/0.181$  & $193.088/193.216/0.162$\\
        Relative energy spread   &   $1.3\times10^{-3}$ &  $1.64\times10^{-3}$ \\
        Damping time(turns) &  80/80/41 & 41/41/21 \\
        Emittance($\text{mm}{\cdot}\text{mrad}$) &  $6.26\times10^{-3}/1.92\times10^{-5}/2.8$ & $9.78\times10^{-3}/2.92\times10^{-5}/4.9$ \\
        Sokolov-Ternov time($\tau_p$)(minute) & 21.4  &  7.0 \\
    Rms spin precession frequency spread($\sigma_{\nu}$) &  0.358  &   0.560   \\
        Modulation index($\alpha$) &  3.921   &   11.986    \\
        Correlation index($\kappa$) &   0.174 &   1.160   \\
        \bottomrule
\end{tabular}
\end{table}

The equilibrium polarization has been computed in three different ways
which are compared with each other. First, the Monte-Carlo approach
described in section \ref{sectionII} has been used to calculate the
equilibrium polarization for $a\gamma_0$ between 267.5 and 275.5 for
Case 1, and between 338.5 and 346.5 for Case 2, the step size for both
cases being $a \gamma_0 = 0.01$. The expected number of emitted photons in a main
dipole is $\sim8$ for 120~GeV and $\sim10$ for 150~GeV, therefore, the
number of integration steps is chosen to be 10 and it is unlikely that
a large number of photons are generated at each integration node.
Sixty particles are tracked for each energy point, and each particle
is tracked for 10 damping times, which takes around 4 minutes for
120~GeV using the CPUs of the Hopper cluster at NERSC.

\begin{figure}[p]
\begin{subfigure}{\linewidth}
\centering
\includegraphics[width=100mm]{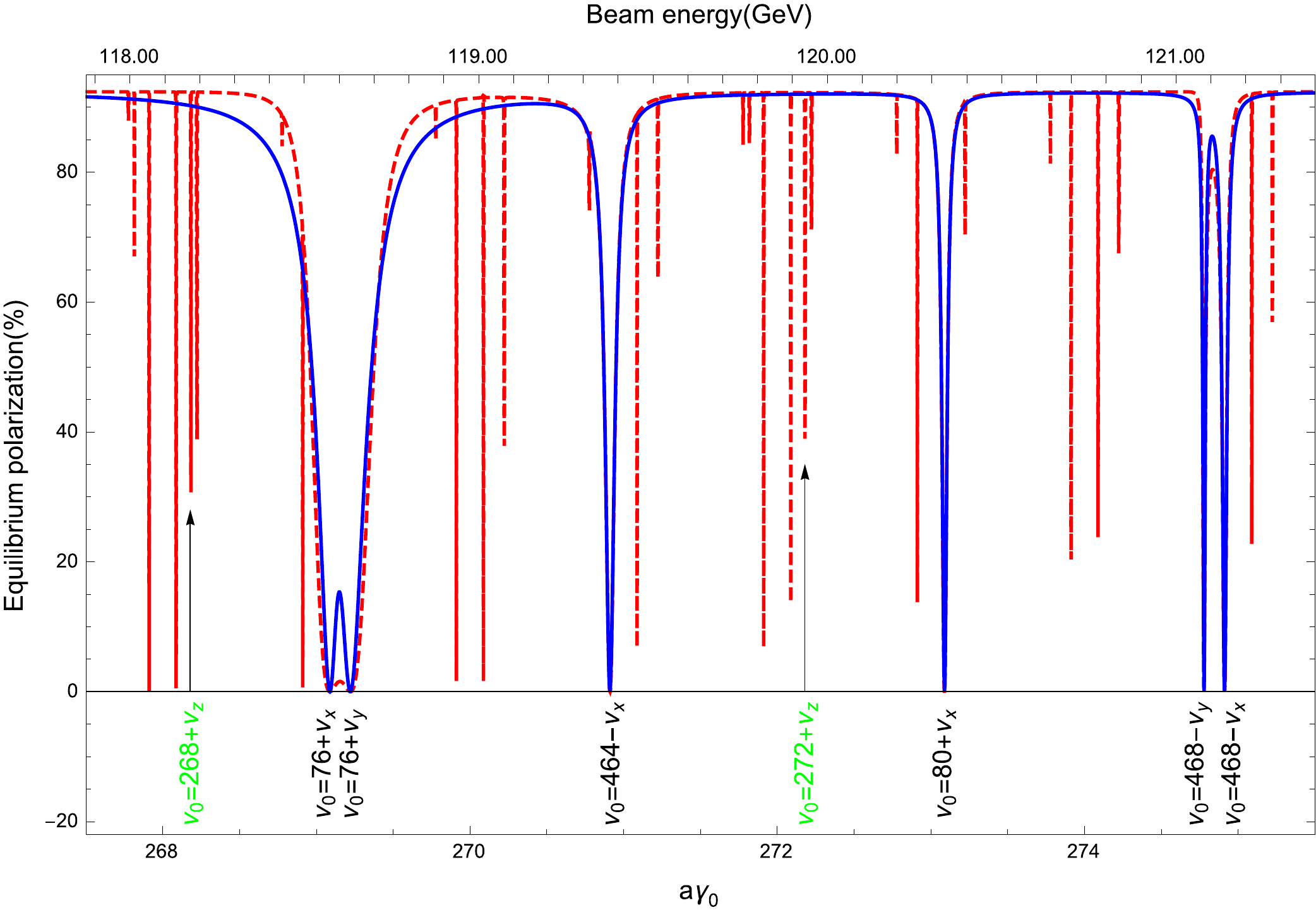}
\caption{Case 1}
\end{subfigure}\\[1ex]
\begin{subfigure}{\linewidth}
\centering
\includegraphics[width=100mm]{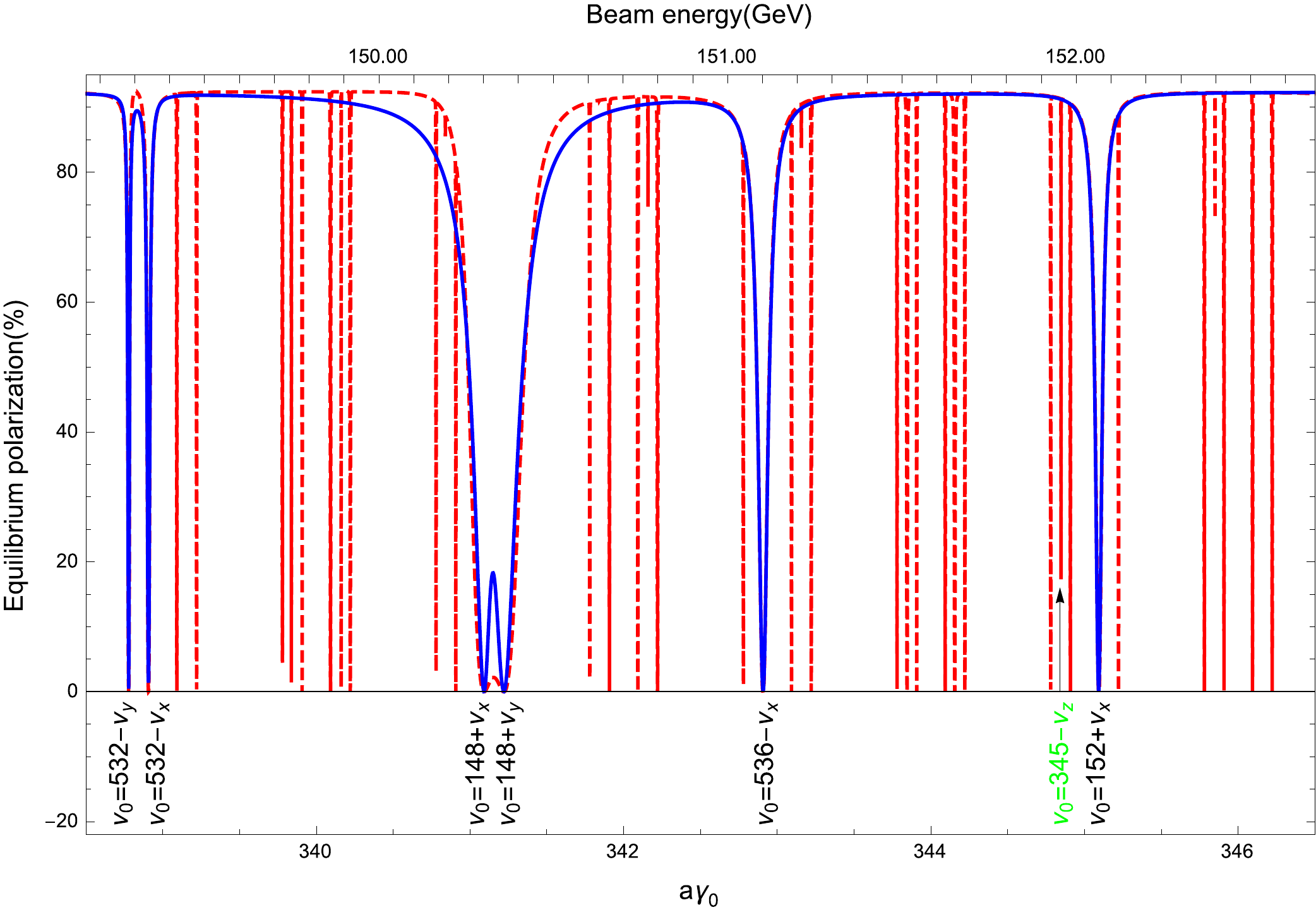}
\caption{Case 2}
\end{subfigure}
 
\caption{Scan of the equilibrium polarization versus beam energy (closed orbit spin tune) with only first-order spin resonances. The dashed line shows the equilibrium polarization computed with a normal form in PTC and the solid line shows the curve fitted using Eq.~\ref{linear bet sideband}. Here, only the betatron spin resonances with $B>10^{-5}$ are retained in the fitting and their locations are indicated at the bottom of the plots. Note that there is no first-order synchrotron spin resonance with $A>10^{-5}$. The locations of some  first-order synchrotron spin resonances are also marked in the plots in green and with arrows.}

\label{linearPol}
\end{figure}

Second, a first order Taylor map is tracked for one turn and $\partial\hat{n}/\partial\delta$
is computed with a normal form \cite{icap2012}. Then the equilibrium polarization
is calculated using Eq.~\ref{DK0} with only the first-order spin resonances.
Following the theory of synchrotron sideband resonances in the
correlated regime, the constants $A$ or $B$ for each first-order 
spin resonance are fitted using Eq.~\ref{linear sync sideband} and Eq.~\ref{linear
bet sideband}. Since the widths of some first-order spin resonances are very small,
and thus their synchrotron sidebands are even narrower, these resonances and
their synchrotron sidebands contribute to the equilibrium beam polarization as
very narrow dips. Since the step size of the Monte-Carlo simulation is
$a\gamma_0=0.01$, some of these very narrow dips are beyond the resolution of the
Monte-Carlo simulation, and thus they are irrelevant for our comparison between
different methods.
As shown in Fig \ref{linearPol}, only the
resonances with $A(B)>10^{-5}$ are retained in the fitting, and no first-order
synchrotron spin resonance qualifies for Case 1 and Case 2. 
Note that for Case 1 near $a\gamma_0=269$, and for Case 2
near $a\gamma_0=341$, the two nearby first-order 
spin resonances overlap with each other, so that there is some
discrepancy between the fitted curve and the simulation result. 
Then the equilibrium polarization
taking into account of the synchrotron sidebands is calculated
following 
Eq.~\ref{beta sideband} and Eq.~\ref{DK1} including 
only the first-order betatron spin resonances with $B>10^{-5}$.

\begin{figure}[p]
\begin{subfigure}{\linewidth}
\centering
\includegraphics[width=100mm]{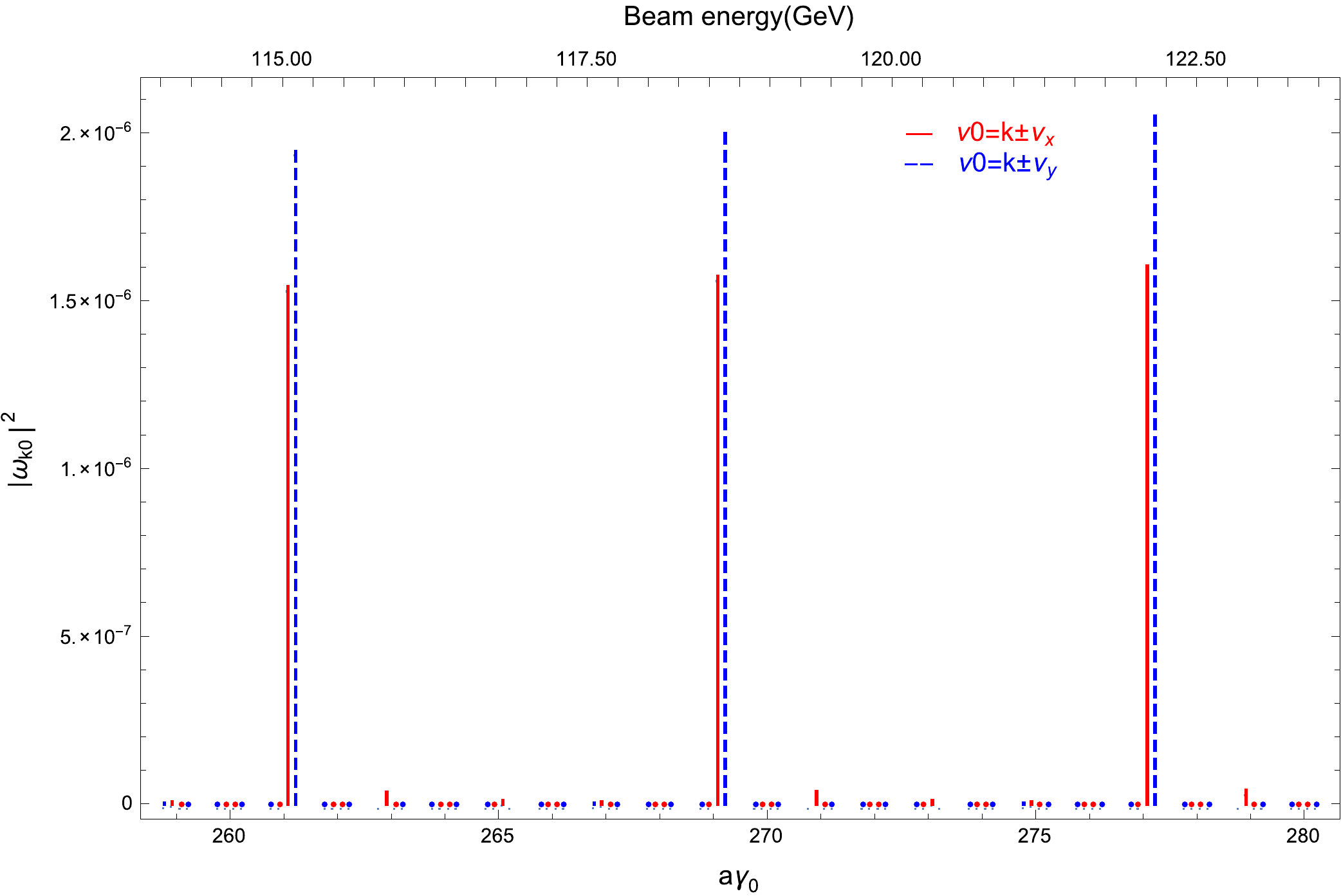}
\caption{Case 1}
\end{subfigure}\\[1ex]
\begin{subfigure}{\linewidth}
\centering
\includegraphics[width=100mm]{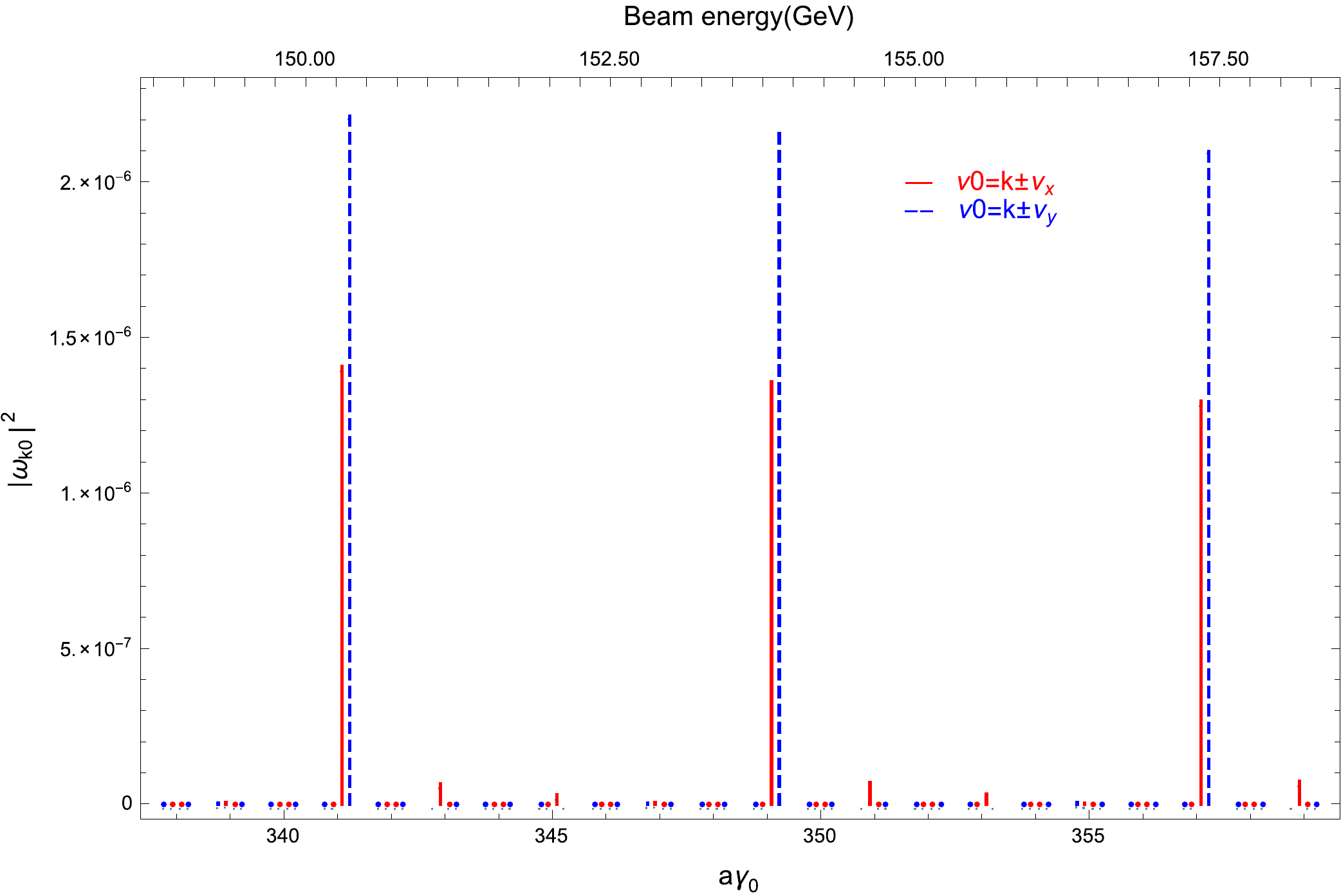}
\caption{Case 2}
\end{subfigure}
\caption{The spectrum of the first-order resonance strengths computed with a normal form in PTC. The resonance strengths are for  particle actions corresponding to the equilibrium beam emittances. The solid and dashed lines represent the horizontal and vertical spin resonances, respectively.}
\label{resonanceStrength}
\end{figure}

Third, following the theory of spin diffusion in the uncorrelated
regime, the strengths of the first-order resonances $\nu_0=k\pm\nu_x$
and $\nu_0=k\pm\nu_y$ are computed with PTC using a normal form 
\cite{icap2012}. For each of these first-order resonances, the
strength scales with the square root of the particle's action, and
when applying Eq.~\ref{deprate} to calculate the depolarization rate
$\lambda'_d$, an average is taken over the equilibrium beam
distribution. Assuming that the beam has a Gaussian distribution, Eq.
\ref{deprate} becomes
\begin{equation}
    \label{deprate1}
    \frac{\lambda'_d} {{\tilde \lambda}_p}=\sum\limits_k\frac{2\sqrt{\pi}\lvert\omega_{k0}\rvert^2} {\sigma_{\nu} {{\tilde \lambda}}_p} e^{-(\nu_0-\nu_k)^2/\sigma_{\nu}^2},
\end{equation}
where $\omega_{k0}$ is the strength of resonance $\nu_0=\nu_k$ of
a particle whose actions correspond to the equilibrium emittances.
It is easy to see from the known dependences of ${{\tilde \lambda}}_p$, $\omega_{k0}$ and ${\sigma}_{\nu}$ on the beam energy 
that ${\lambda'_d}/ {{\tilde \lambda}_p}$ decreases strongly as the beam energy increases.
The  $\omega_{k0}$ of
the first-order resonances in the two energy ranges are shown in Figure \ref{resonanceStrength}. 
Since it is not clear to us whether $\lambda^0_d$ should be added
to the depolarization rate as well, we just ignore its
contribution in this method. 

\begin{figure}[p]
\begin{subfigure}{\linewidth}
\centering
\includegraphics[width=100mm]{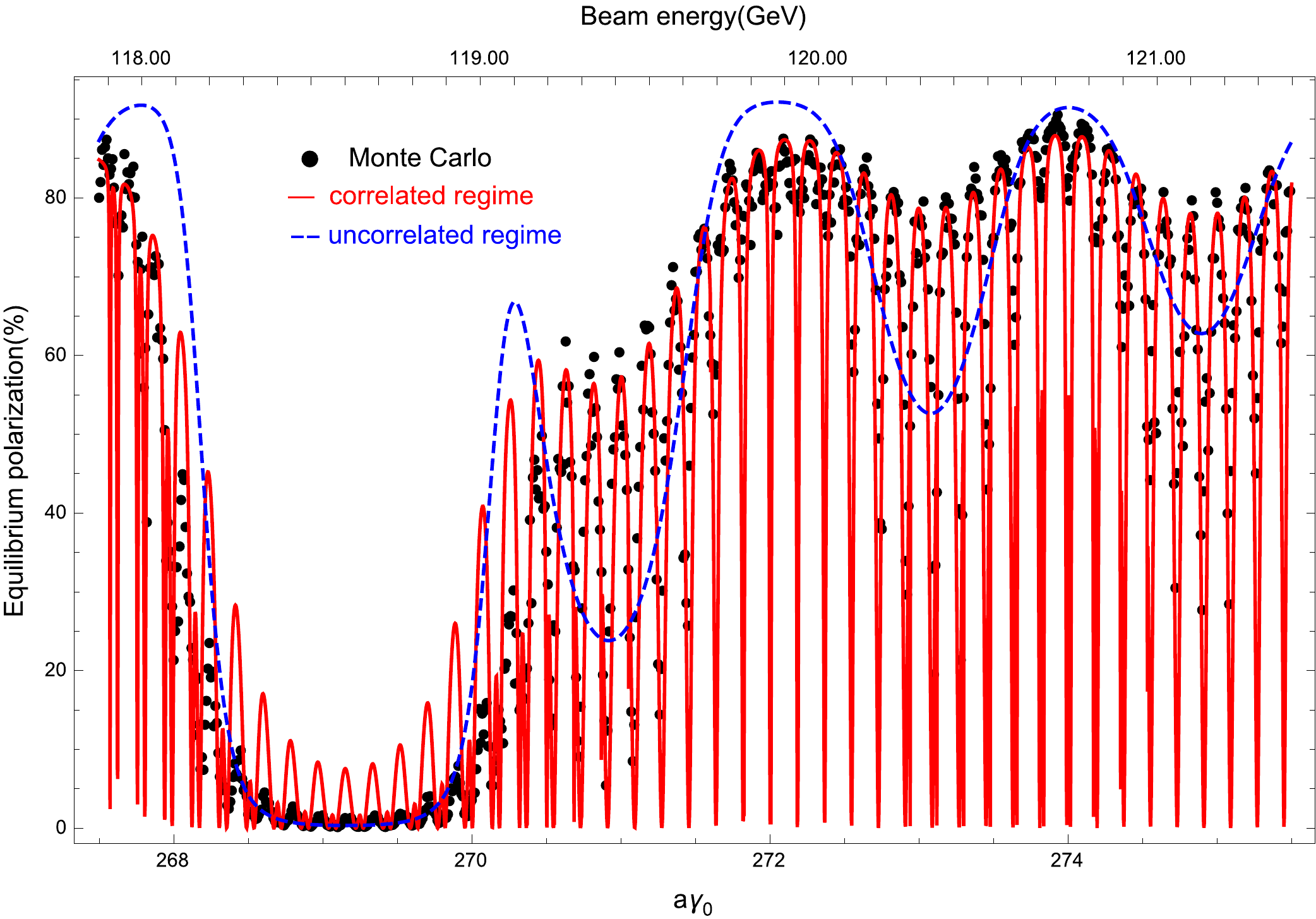}
\caption{Case 1}
\end{subfigure}\\[1ex]
\begin{subfigure}{\linewidth}
\centering
\includegraphics[width=100mm]{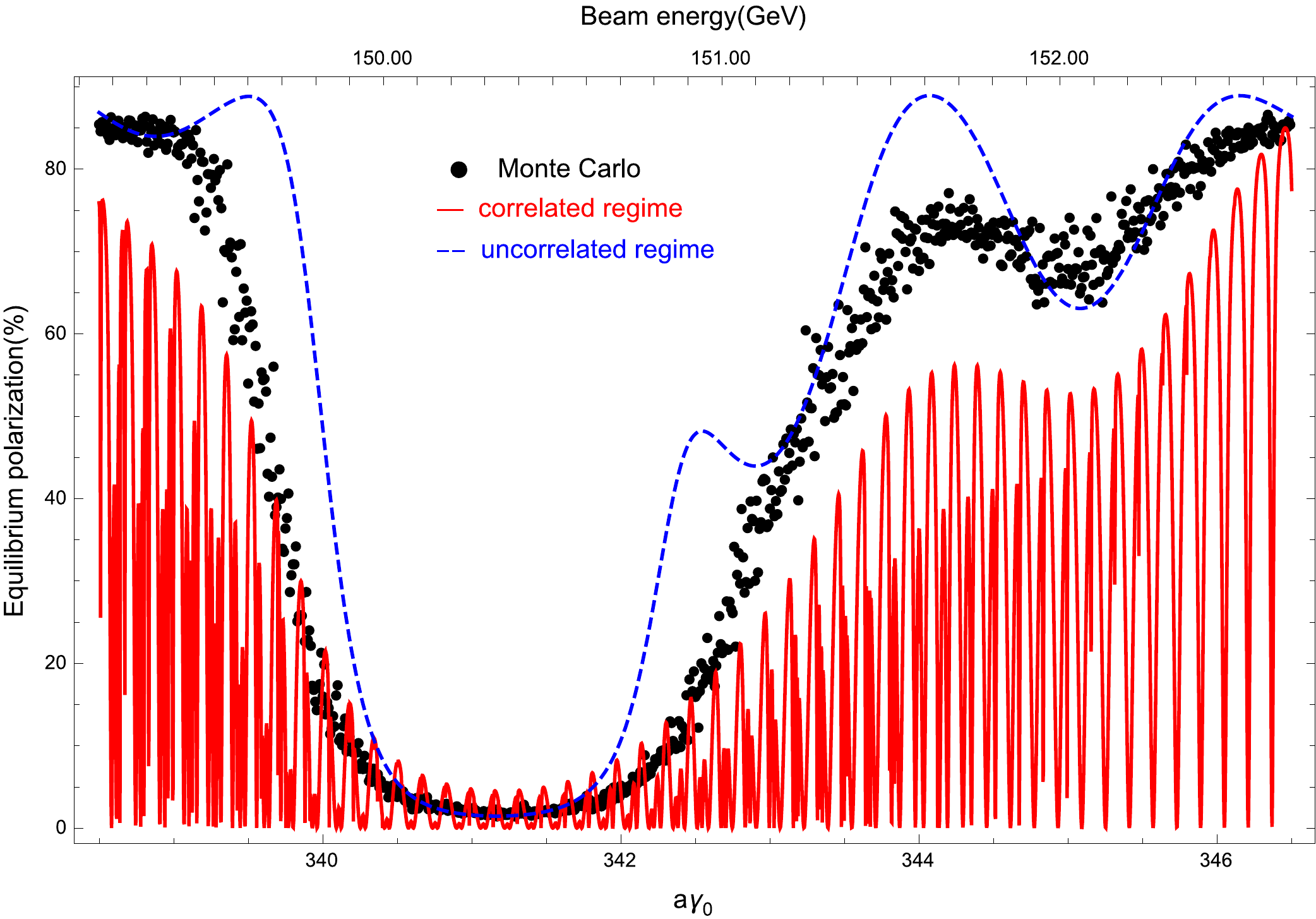}
\caption{Case 2}
\end{subfigure}
\caption{Comparison of the equilibrium polarizations computed by the three different methods. The first method is
the Monte-Carlo simulation. The second method first fits the widths of parent spin resonances using Eq.~\ref{linear bet sideband}, and then computes the 
synchrotron sideband resonances following Eq.~\ref{beta sideband},
and it is labeled  as ''correlated regime'' in this figure. The third method, referred to as ''uncorrelated regime'' in this figure, computes the strengths of nearby first-order spin resonances using a normal form, and then applies Eq.~\ref{deprate1}.}
\label{Poltotal}
\end{figure}

Results for the three methods are shown in Figure
\ref{Poltotal} for both cases. For Case 1, the correlation index 
is $0.174$, much smaller than 1, and the Monte-Carlo result is
more consistent with the expectation for correlated regime, where fine resonance
structures are observed. For Case 2, the correlation index is
$1.160$, and according to the criterion,
it is regarded as being in the uncorrelated regime.
The Monte-Carlo result does not show 
clear synchrotron-sideband structures, and it looks more
consistent with that of the uncorrelated regime. The discrepancy between
the result of the Monte-Carlo simulation and that for the ''uncorrelated regime'' might
be due to the ignored contribution from $\lambda^0_d$. This needs
further verification and that requires a code to evaluate the spin
orbit coupling function non-perturbatively. SODOM and stroboscopic averaging
are two possible options, but their application in such ultra-high beam energies
also needs justification. This is beyond the scope of this paper.

In any case, so far, and without misalignments, the results of the simulations support the theory of the
uncorrelated resonance crossing in synchrotron oscillations. In
addition, it also indicates Eq.~\ref{DK0} is not applicable for
evaluating the equilibrium beam polarization in the uncorrelated
regime.

\section{\label{sectionIV} Conclusion}
This paper presents Monte-Carlo simulations of the equilibrium beam
polarization in electron storage rings, on the basis of the PTC
code, which treats nonlinear orbit and three-dimensional spin motion, and which generates synchrotron-radiation 
photons at each integration step. 
The simulations  are
benchmarked against SODOM \cite{Yokoya:1992nt}, and limitations
are also discussed. 
Monte-Carlo simulations are launched to study a model storage ring with parameters similar to those of the proposed CEPC, and the
results are consistent with the suggestions for the correlated and uncorrelated regimes of beam depolarization at ultra-high beam
energies.

In particular, the results are consistent with the notion 
that already at 150~GeV more polarization is attainable than that predicted by Eq.
\ref {DK0} with the analytical solution for synchrotron sideband resonances.
This is obviously very important. Nevertheless, we feel that
the notion of uncorrelated resonance crossing should be put on a
firmer mathematical basis. A good starting point would be 
rigorous solution of the evolution equation for the polarization density \cite{dbkh98}.
In any case, the simulations suggest that
use of a Monte-Carlo method is the most effective way to proceed at present. As
far as we are aware this is the first serious numerical study of
this kind for ultra-high beam energies. 

At this stage of our
studies our models do not have misalignments and the resulting
distortions of the closed orbit. However, the simulation of
realistic misalignments and the modeling of the correction of the
orbital imperfections are very important for beam polarization at
ultra-high beam energies. These topics will be studied in the future
on the basis of this simulation framework,
and will be an integral part of investigation of feasibility
of attaining high beam polarization at CEPC. Nevertheless our current
results are not just of academic interest.

For the attainment of polarized beam at not-so-ultra-high beam
energies (for example 120~GeV), it appears to be necessary to find
some way to reduce the spread in the spin-precession rate. It has been
suggested that this could be achieved by 
manipulating the dispersion functions \cite{Yokoya_spread}
or by reducing the energy
spread with the aid of special nonlinear wigglers 
\cite{Jowett1, Jowett2}. However, it is far from clear that such
schemes are practical. An alternative would be to introduce two
Siberian snakes. For this, ${\hat n}_0$ points upwards in one half ring
and it points downwards in the other half so that Sokolov-Ternov
effect is suppressed. This could be overcome by having short
strong dipoles in one half ring and weak long dipoles in the other
half. A simulation on this at lower energies 
\cite{Barberspin2010} has already shown a suppression of synchrotron
sideband resonances.

\appendix
This work is supported by the Hundred-Talent Program (Chinese Academy
of Sciences), and National Natural Science Foundation of China
(11105164). We would like to thank Drs. D. Abell and E. Forest on
their help with the simulation code PTC. One of us, Z. Duan, would
like to thank Drs. S. Mane, U. Wienands and K. Yokoya for helpful
discussions in theoretical aspects. The simulation work used the
resources of the National Energy Research Scientific Computing
Center, a DOE Office of Science User Facility supported by the Office
of Science of the U.S. Department of Energy under Contract No.
DE-AC02-05CH11231.


\begin{thebibliography}{10}
\expandafter\ifx\csname url\endcsname\relax
  \def\url#1{\texttt{#1}}\fi
\expandafter\ifx\csname urlprefix\endcsname\relax\def\urlprefix{URL }\fi
\expandafter\ifx\csname href\endcsname\relax
  \def\href#1#2{#2} \def\path#1{#1}\fi

\bibitem{Schmidt:2002vp}
F.~Schmidt, E.~Forest, E.~McIntosh, CERN-SL-2002-044-AP, KEK-REPORT-2002-3 (2002).

\bibitem{Derbenev:1979tm}
Y.~Derbenev, A.~Kondratenko, A.~Skrinsky, Part. Accel. 9 (1979) 247.

\bibitem{CEPC}
{CEPC}, {http://cepc.ihep.ac.cn}.

\bibitem{FCC-ee}
{FCC-ee}, {http://tlep.web.cern.ch}.

\bibitem{L.H.Thomas1927}
L.~H.Thomas, Phil.Mag. 3 (1927) 1-21.

\bibitem{V.Bargmann1959}
V.~Bargmann, L.~Michel, V.~L. Telegdi, Phys. Rev. Lett. 2 (1959) 435.

\bibitem{D.P.Barber2004}
D.~P.~Barber, J.~Ellison, K.~Heinemann, Phys.Rev. ST Accel.Beams 7 (2004) 124002.

\bibitem{Sokolov:1963zn}
A.~Sokolov, I.~Ternov, Sov. Phys. Doklady 8 (1964) 1203.

\bibitem{V.N.Baier1966}
V.~N. Baier, Y.~Orlov, Sov. Phys. Doklady 10 (1966) 1145.

\bibitem{Derbenev:1973ia}
Y.~Derbenev, A.~Kondratenko, Sov. Phys. JETP 37 (1973) 968--973.

\bibitem{Chao:1999qt}
D.~P. ~Barber, G.~Ripken, in: {Handbook of Accelerator Physics and Engineering},
A. W. Chao, M. Tigner(eds.), 1st ed., 3rd printing, World Scientific, Singapore (2006); 
A. W. Chao, K.H. Mess, M. Tigner, F. Zimmermann(eds.), 2nd ed., World Scientific, Hackensack, NJ, USA, (2013).

\bibitem{Chao:1980fz}
A.~Chao, Nucl. Instrum. Meth. 180 (1981) 29.

\bibitem{Mane:1986qz}
S.~Mane, Phys. Rev. A 36 (1987) 120.

\bibitem{icap2012}
D.~.T. Abell, D.~Meiser, D.~P. Barber and E.~Forest, Talk on ICAP 2012, Rostock-Warnemünde, Germany, 2012.

\bibitem{S.R.ManePrivate}
S.~Mane, {Private communication}.

\bibitem{Yokoya:1992nt}
K.~Yokoya, KEK Report KEK-92-6, KEK (1992).

\bibitem{K.Heinemann1996}
K.~Heinemann, G.~H. Hoffst{\"a}tter, Physical Review E 54 (1996) 4240.

\bibitem{Kewisch:1983uq}
J.~Kewisch, DESY Report DESY-83-032, DESY (1983).

\bibitem{Kewisch:1988nv}
J.~Kewisch, R.~Rossmanith, T.~Limberg, Phys. Rev. Lett. 62 (1989) 419.

\bibitem{M.Boge}
M.~Boge, DESY report, DESY-94-087, DESY (1994).

\bibitem{M.Berglund}
M.~Berglund, Doctoral thesis, Royal Institute of Technology, Stockholm, Sweden (2001); 
 Report DESY-Thesis-2001-044 (2001).

\bibitem{Barber:2005slicktrack}
D.~P. Barber, in: Proc. 16th Int. Spin Physics Symp., World Scientific, Kyoto, Japan, 2005.

\bibitem{S.Mane2009PTC}
S.~Mane, KEK Report KEK-2009-8, KEK (2009).

\bibitem{fppipac2006}
E.~Forest, Y.~Nogiwa, in: {Proc. ICAP 2006},  Chamonix, France, 2006.

\bibitem{MADX}
{Methodical Accelerator Design (MAD) program}, {CERN, http://cern.ch/mad/}.

\bibitem{Sagan:2006sy}
D.~Sagan, Nucl. Instrum. Meth. A 558 (2006) 356.

\bibitem{Cai:1998zp}
Y.~Cai, SLAC-PUB-7793, SLAC (1998).

\bibitem{Roy:1990vw}
G.~Roy, Nucl. Instrum. Meth. A 298 (1990) 128.

\bibitem{Burkhardt:2007zza}
H.~Burkhardt, EUROTEV-REPORT-2007-018, CLIC-NOTE-709, CERN-OPEN-2007-018, CERN (2007).

\bibitem{Alexahin:2014cha}
Y.~Alexahin, arXiv:1409.5483 (2014).

\bibitem{ohmienv}
K.~Ohmi, K.~Hirata, K.~Oide, Phys. Rev. E. 49 (1994) 751.

\bibitem{S.R.Mane2005}
S.~Mane, Y.~M. Shatunov, K.~Yokoya, Rep. Prog. Phys. 68 (2005) 1997.

\bibitem{Barber:1994ew}
D.~Barber, et~al., Phys. Lett. B343 (1995) 436.

\bibitem{Froissart1960}
M. Froissart and R. Stora,  Nucl. Instrum. Meth. 7 (1960) 297.

\bibitem{Hoff's book}
G. H. Hoffstaetter, ''High-energy polarized proton beams: a modern view, in: Springer Tracts in Modern Physics'', Volumn 218, 2006.

\bibitem{Derbenev1977}
Y.~Derbenev, A.~Kondratenko, in: {Proc. 10th Int. Conf. on High Energy
  Accelerators, Protvino, USSR}, 1977.

\bibitem{Derbenev1978}
Y.~Derbenev, A.~Kondratenko, {Part. Accel.} 8 (1978) 115.

\bibitem{PTClibug}
    D. Abell, {PTC Library User Guide}, Tech-X Corporation, unpublished, 2011.

\bibitem{Barberspin2010}
D.P. Barber, H-U. Wienands, M. Fitterer and  H. Burkhardt,
in: {Proc. 19th Int. Spin Physics Symposium, Juelich, Germany}, 2010.


\bibitem{Mane:1990ge}
S.~Mane, Nucl. Instrum. Meth. A292 (1990) 52.

\bibitem{Derbenev1975}
Y.~Derbenev, A.~Kondratenko,  Sov. Phys. Doklady 19 (1975) 438.


\bibitem{Kondratenko1974}
A.~Kondratenko,  Sov. Phys. Doklady 39 (1974) 592.

\bibitem{Yokoya:1983sm}
K.~Yokoya, Part. Accel. 14 (1983) 39.

\bibitem{Sands:1970ye}
M.~Sands, SLAC-R-121, SLAC (1970). 

\bibitem{Assmann}
R. Assmann, in: Proc. PAC 1999, New York City, USA, 1999.

\bibitem{101/45optics}
R. Assmann, et~al., CERN-SL-2000-012, CERN (2000).

\bibitem{dbkh98}
K. Heinemann and D.P. Barber, Nucl. Instr. Meth. A {\bf 463} (2001) 62
and {\bf 469} (2001) 294.

\bibitem{Yokoya_spread}
K. Yokoya, Part. Accel. 13 (1983) 95.

\bibitem{Jowett1}
A. Hofmann and J. M. Jowett, CERN/ISF-TH/81-23, CERN (1981). 

\bibitem{Jowett2}
J. M. Jowett, CERN/ISF-TH/81-24, CERN (1981).

\end{thebibliography}
\end{document}